\documentclass[12pt]{article}
\usepackage{axodraw}
\usepackage{amsmath}
\usepackage{graphics}
\newcommand{\as}{\alpha_{\mathrm{s}}}
\newcommand{\aem}{\alpha_{\mathrm{em}}}
\renewcommand{\d}{\mathrm{d}}
\renewcommand{\u}{\mathrm{u}}
\newcommand{\s}{\mathrm{s}}
\renewcommand{\c}{\mathrm{c}}

\newcommand{\X}{\mathbf{X}}
\newcommand{\e}{\mathrm{e}}

\newcommand{\g}{\mathrm{g}}
\newcommand{\p}{\mathrm{p}}
\newcommand{\q}{\mathrm{q}}

\newcommand{\qbar}{\mathrm{\overline{q}}}

\newcommand{\pT}{p_{\perp}}
\newcommand{\kT}{k_{\perp}}
\newcommand{\kTone}{k_{\perp 1}}
\newcommand{\kTtwo}{k_{\perp 2}}
\newcommand{\pTmin}{p_{\perp\mathrm{min}}}

\newcommand{\gast}{\gamma^*}
\newcommand{\ga}{\gamma}

\newcommand{\Jpsi}{\mathrm{J}/\psi}
\newcommand{\mr}{\mathrm}
\newcommand{\ra}{\rightarrow}
\newcommand{\lessim}{\raisebox{-0.8mm}%
{\hspace{1mm}$\stackrel{<}{\sim}$\hspace{1mm}}}
\def\Journal#1#2#3#4{{#1}{\bf #2} (#3) #4}

\def\NPB{{\em Nucl. Phys.~}{\bf B}}
\def\PLB{{\em Phys. Lett.~}{\bf B}}

\def\PRL{\em Phys. Rev. Lett.~}
\def\PRD{{\em Phys. Rev.~}{\bf D}}
\def\ZPC{{\em Z. Phys.~}{\bf C}}

\def\JHEP{\em J. High Energy Phys.~}
\def\CPC{\em Computer Phys. Commun.~}
\def\PRP{\em Phys. Rept.~}
\def\PRV{\em Phys. Rev.~}
\def\EPJC{{\em Eur. Phys. J.~}{\bf C}}
\def\RMP{\em Rev. Mod. Phys.~}
\def\SJNP{\em Sov. J. Nucl. Phys.~}
\def\SPJP{\em Sov. Phys. JETP~}
\def\YF{\em Yad. Fiz.~}
\def\ZETF{\em Zh. Eksp. Teor. Fiz.~}
\newenvironment{Itemize}{\begin{list}{$\bullet$}%
{\setlength{\topsep}{0.2mm}\setlength{\partopsep}{0.2mm}%
\setlength{\itemsep}{0.2mm}\setlength{\parsep}{0.2mm}}}%
{\end{list}}
\newcounter{enumct}
\newenvironment{Enumerate}{\begin{list}{\arabic{enumct}.}%
{\usecounter{enumct}\setlength{\topsep}{0.2mm}%
\setlength{\partopsep}{0.2mm}\setlength{\itemsep}{0.2mm}%
\setlength{\parsep}{0.2mm}}}{\end{list}}
\newlength{\abstwidth}
\newlength{\captivewidth}
\begin{document}
%
 
\sloppy
 
\pagestyle{empty}
 
\begin{flushright}
LU TP 00--15\\
hep-ph/0005048\\
May 2000
\end{flushright}
 
\vspace{\fill}
 
\begin{center}
{\LARGE\bf Total Cross Sections with Virtual Photons%
\footnote{To appear in the Proceedings of the LEP2 Workshop on Monte Carlo generators
}}\\[10mm]
{\Large Christer Friberg\footnote{christer@thep.lu.se}}\\[2mm]
{\it Department of Theoretical Physics,}\\[1mm]
{\it Lund University, Lund, Sweden}
\end{center}
 
\vspace{\fill}
\begin{center}
{\bf Abstract}\\[2ex]
\begin{minipage}{\abstwidth}
A model for total cross sections with virtual photons is presented, 
in particular $\gast\p$ and $\gast\gast$ cross sections are 
considered. Our approach is based on an existing model for 
photoproduction, which subdivides the total cross section into 
three distinct event classes: direct, VMD and 
anomalous \cite{SchSjgap}. In the region of large photon 
virtualities, the Deep Inelastic Scattering processes 
(up to $\mathcal{O}(\alpha_s)$ corrections) are obtained. 
Hence, the model provides a smooth transition between the 
two regions. By the breakdown into different event classes, a
complete picture of all event properties is the ultimate goal.
\end{minipage}
\end{center}

\vspace{\fill}

\clearpage
\pagestyle{plain}
\setcounter{page}{1}

\section{Introduction}

{\sc Pythia} is a general-purpose event generator of high-energy particle 
physics reactions. A full description of models and processes implemented
are found in Ref.~\cite{pythia} with relevant update notes on the 
{\sc Pythia} webpage. Some aspects specific to $\ga\ga$ physics are 
summarized in Ref.~\cite{YR96py} and will briefly be reviewed here 
to illustrate the modifications and improvements done to also treat 
virtual photons.
The model used so far, incorporating Leading Order (LO) hard 
scattering processes, as well as elastic, diffractive, low-$\pT$ and 
multiple parton--parton scattering for specific event classes, considers 
only real incoming photons with
a separate treatment of the Deep Inelastic Scattering (DIS) region, 
$\e\ga \ra \e\X$~\cite{SchSjgap}. 
Here, the two extreme scenarios will be merged in order to obtain a 
description that smoothly interpolates between the two regions.  

A model for jet production with virtual photons has been described in detail
elsewhere~\cite{lutp9911}. Photon flux factors are convoluted with 
matrix elements involving either direct 
or resolved photons and, for the latter, with parton distributions 
of the photon. The direct and single-resolved matrix elements are depending
on the virtuality of the photon and the virtual resolved photons are 
dampened with dipole factors in the parton distributions.
The range of uncertainty in the modeling of the resolved component was 
explored, eg. parton distribution sets of the photon, scale choice
in the parton distributions, longitudinal contributions etc.

In this report we will limit ourself to the discussion of mixing different event classes to obtain total cross sections with virtual photons, 
$\gast\p$ and $\gast\gast$. The extension to virtual photons will eventually 
also be made for elastic scattering and diffractive events. This will 
be described in a future publication~\cite{future} together with a 
more detailed description of the model presented here. 

\section{Event Classes}

Traditionally, different descriptions are used for virtual and real
photons. Virtual photons in the DIS  
region are normally described as devoid of any structure, while for 
the real ones, the possibility of hadronic-like fluctuations play an 
important r\^ole. In the region of intermediate $Q^2$, it should be 
possible to find a description starting from either extreme. Then the 
language may not always be unique, i.e. a given Feynman diagram may be 
classified in different ways. In the following, we will develop one 
specific approach, where the main idea is to classify events by the 
hardest scale involved. 

We begin by a reminder on the models for photoproduction 
and DIS, before embarking on the generalization also to  
intermediate virtualities in $\ga^*\p$ processes. The $\ga\ga$, 
$\ga^*\ga$ and $\ga^*\ga^*$ processes thereafter follow by 
an application of the same rules.

\subsection{Photoproduction}

To first approximation, the photon is a point-like particle. However,
quantum mechanically, it may fluctuate into a (charged) 
fermion--antifermion pair. The fluctuations 
$\ga \leftrightarrow \q\qbar$ can interact strongly and therefore 
turn out to be responsible for the major part of the $\ga\p$ and 
$\ga\ga$ total 
cross sections. The total rate of $\q\qbar$ fluctuations is not 
perturbatively calculable, since low-virtuality fluctuations enter a 
domain of non-perturbative QCD physics. It is therefore customary to split
the spectrum of fluctuations into a low-virtuality and a high-virtuality
part. The former part can be approximated by a sum over low-mass 
vector-meson states, customarily (but not necessarily) restricted 
to the lowest-lying vector multiplet. Phenomenologically, this 
Vector Meson Dominance (VMD) ansatz turns out to be very successful in
describing a host of data. The high-virtuality part, on the other hand, 
should be in a perturbatively calculable domain. Based on the above 
separation, three main classes of interacting photons can be 
distinguished: direct, VMD and anomalous photons, corresponding to the 
following event classes in $\ga\p$ events 
\cite{SchSjgap}:
\begin{Enumerate}
\item The VMD processes, where the photon turns into a vector meson
before the interaction, and therefore all processes
allowed in hadronic physics may occur. This includes elastic and 
diffractive scattering as well as low-$\pT$ and high-$\pT$ 
non-diffractive events.
\item The direct processes, where a bare photon interacts with a 
parton from the proton.
\item The anomalous processes, where the photon perturbatively branches
into a $\q\qbar$ pair, and one of these (or a daughter parton thereof)
interacts with a parton from the proton. 
\end{Enumerate}
The total photoproduction cross section can then be written as
$
\sigma_{\mr{tot}}^{\ga\p}=
\sigma_{\mr{VMD}}^{\ga\p}+
\sigma_{\mr{direct}}^{\ga\p}+
\sigma_{\mr{anomalous}}^{\ga\p} \;.
\label{eq:evclassgp}
$

Total hadronic cross sections show a characteristic fall-off at 
low energies and a slow rise at higher energies. This behaviour 
can be parameterized by the form 
\begin{equation}
\sigma_{\mr{tot}}^{AB}(s) = X^{AB} s^{\epsilon} + Y^{AB} s^{-\eta}
\label{sigmatotAB}
\end{equation}
for $A + B \to X$. The powers $\epsilon$ and $\eta$
are universal, with fit values \cite{DL92}
$
  \epsilon \approx 0.0808, 
  \eta \approx 0.4525,
$
while the coefficients $X^{AB}$ and $Y^{AB}$ are
process-dependent. Equation (\ref{sigmatotAB}) can be interpreted 
within Regge theory, 
but for the purpose of our study we 
can merely consider it as 
a convenient parameterization.

The VMD part of the $\ga\p$ cross section is an obvious candidate
for a hadronic description. The diagonal VMD model suggests:
\begin{equation}
\sigma_{\mr{VMD}}^{\ga\p}(s) = 
\sum_{V=\rho^0,\omega,\phi,\Jpsi}\; 
\frac{4\pi\aem}{f_V^2}\;
\sigma_{\mr{tot}}^{V\p}(s) 
\approx  53.4 s^{\epsilon} + 115 s^{-\eta}~~[\mu\mr{b}]\;,
\label{sigmatotVMDp}
\end{equation} 
with the $f_V$ determined from data~\cite{Baur}. 
The $V\p$ cross sections can be parameterized 
assuming an additive quark model \cite{SchSjgap} 
and adding the vector meson contributions, we arrive at the above 
numbers (with $s$ in GeV$^2$).

There is no compelling reason that such an ansatz should hold also for the 
total $\ga\p$ cross section, but empirically a parameterization according to
eq.~(\ref{sigmatotAB}) does a good job \cite{DL92}. For instance, 
such parameterization predicted the high-energy behaviour of the cross 
section, close to what was then measured by H1 and ZEUS. 
Thus VMD corresponds to approximately 80~\% of the total $\ga\p$ cross 
section at high energies, with the remaining 20~\% then 
shared among the direct and anomalous event classes.

Introducing a cut-off parameter $k_0$ to separate the low- and 
high-virtuality parts of the $\q\qbar$ fluctuations, the anomalous 
contribution can be written as
\begin{equation}
\sigma_{\mr{ano}}^{\ga\p}(s)=
\frac{\aem}{2\pi} \; \sum_\q 2 \e_\q^2 \int_{k_0^2}^{\infty} 
\frac{\d \kT^2}{\kT^2} \;
\sigma^{\q\qbar\p}(s; \kT)
=
\frac{\aem}{2\pi} \; \sum_\q 2 \e_\q^2 \int_{k_0^2}^{\infty} 
\frac{\d \kT^2}{\kT^2} \;
\frac{k_{V(\q\qbar)}^2}{\kT^2} \;
\sigma^{V(\q\qbar)\p}(s)
\label{eq:anoint}
\end{equation}
where the prefactor and integral over $\d\kT^2/\kT^2$ corresponds to the 
probability for the photon to split into a $\q\qbar$ state of transverse 
momenta $\pm \kT$. The cross section for this $\q\qbar$ pair to scatter 
against the proton, $\sigma^{\q\qbar\p}$, need to be modeled. 
Based only on geometrical scaling arguments, one could expect a decrease 
roughly like $1/\kT^2$ which motivates the second equality.
The $k_{V(\q\qbar)}$ is a free parameter introduced for dimensional 
reasons. It could be associated with the typical $\kT$ inside the
vector meson $V$ formed from a $\q\qbar$ pair: $\rho^0 \approx \omega$ 
for $\u$ and $\d$, $\phi$ for $\s$, $\Jpsi$ for $\c$. As a reasonable 
ansatz, one could guess $k_{V(\q\qbar)} \approx m_V/2 \approx m_{\rho}/2$.  
Fits to the total cross section at not too
high energies, with a large VMD and a small direct contributions subtracted, 
give corresponding numbers, $k_{V(\q\qbar)} \approx 0.41$~GeV for a
$k_0 \approx 0.5$~GeV.  

To leading order, the direct events come in two kinds: QCD Compton  
$\ga \q \to \q \g$ (QCDC) and boson-gluon fusion 
$\ga \g \to \q \qbar$ (BGF). The cross sections are divergent in 
the limit $\kT \to 0$ for the outgoing parton pair. Therefore 
a lower cut-off is required, but no other specific model assumptions. 

The subdivision of the photon into three different components leads to
the existence of three times three event classes in $\ga\ga$ events. 
By symmetry, the `off-diagonal' combinations appear
pairwise, so for real photons the number of distinct classes is only 
six. These are, 
\begin{Enumerate}
\item VMD$\times$VMD: both photons turn into hadrons, and the processes
are therefore the same as allowed in hadron--hadron collisions.
\item VMD$\times$direct: a bare photon interacts with the partons of the
VMD photon.
\item VMD$\times$anomalous: the anomalous photon perturbatively 
branches into a $\q\qbar$ pair, and one of these (or a daughter parton 
thereof) interacts with a parton from the VMD photon.
\item Direct$\times$direct: the two photons directly give a quark pair,
$\ga\ga \to \q\qbar$. Also lepton pair production is allowed,
$\ga\ga \to \ell^+\ell^-$, but will not be considered by us.
\item Direct$\times$anomalous: the anomalous photon perturbatively 
branches into a $\q\qbar$ pair, and one of these (or a daughter parton 
thereof) directly interacts with the other photon. 
\item Anomalous$\times$anomalous: both photons perturbatively branch 
into $\q\qbar$ pairs, and subsequently one parton from each photon 
undergoes a hard interaction.
\end{Enumerate}
Most of the above classes above are pretty much the same as  allowed in 
$\ga\p$ events, since the interactions of a VMD or anomalous photon and 
those of a proton are about the same. Only the direct$\times$direct
class offer a new hard subprocess.

The main parton-level processes that occur in the above classes are:
\begin{Itemize}
\item The `direct' processes $\ga\ga \to \q\qbar$ only occur 
in class 4.
\item The `1-resolved' processes $\ga\q \to \q\g$ and 
$\ga\g \to \q\qbar$ occur in classes 2 and 5.
\item The `2-resolved' processes $\q\q' \to \q\q'$ (where $\q'$ 
may also represent an antiquark), $\q\qbar \to \q'\qbar'$,
$\q\qbar \to \g\g$, $\q\g \to \q\g$, $\g\g \to \q\qbar$ and
$\g\g \to \g\g$ occur in classes 1, 3 and 6.
\end{Itemize} 

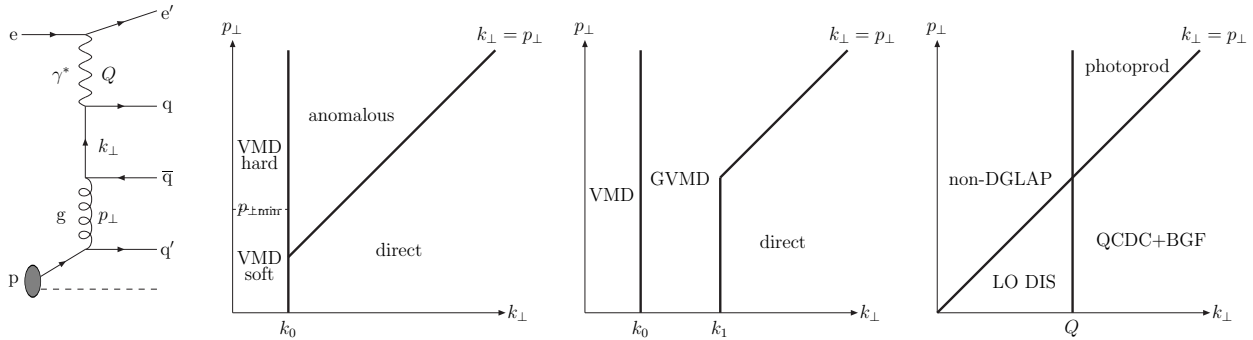
\begin{figure}[t]
\begin{center}  
\scalebox{0.6}{
\begin{picture}(105,220)(-4,-25)
  \ArrowLine(5,165)(45,165)
  \ArrowLine(45,165)(90,180)
  \Photon(45,165)(45,120){4}{4}  
  \GOval(12,10)(10,5)(0){0.5}
  \DashLine(17,5)(90,5){4}
  \ArrowLine(17,13)(45,30)
  \Gluon(45,75)(45,30){4}{4}
  \ArrowLine(45,75)(45,120)
  \ArrowLine(45,30)(90,30)
  \ArrowLine(90,75)(45,75)
  \ArrowLine(45,120)(90,120) 
  \Text(0,10)[]{$\p$}
  \Text(0,165)[]{$\e$}
  \Text(30,140)[]{$\ga^*$}     
  \Text(30,50)[]{$\g$}     
  \Text(60,50)[]{$\pT$} 
  \Text(60,95)[]{$\kT$} 
  \Text(60,140)[]{$Q$} 
  \Text(97,30)[]{$\q'$} 
  \Text(97,75)[]{$\qbar$} 
  \Text(97,120)[]{$\q$} 
  \Text(97,180)[]{$\e'$} 
\end{picture} 
\hspace{5mm}
\begin{picture}(200,200)(0,0)
  \LongArrow(15,15)(185,15)
  \Text(195,15)[]{$\kT$} 
  \LongArrow(15,15)(15,185) 
  \Text(15,194)[]{$\pT$} 
  \SetWidth{1.5}
  \Line(50,15)(50,180)
  \Text(50,5)[]{$k_0$}
  \Line(50,50)(180,180)
  \Text(190,190)[]{$\kT = \pT$}
  \SetWidth{0.3}
  \DashLine(15,80)(50,80){2}
  \SetWidth{0.5}
  \Text(32,120)[]{VMD}
  \Text(32,110)[]{hard}   
  \Text(32,80)[]{$\pTmin$}   
  \Text(32,50)[]{VMD}
  \Text(32,40)[]{soft}
  \Text(120,55)[]{direct}
  \Text(90,140)[]{anomalous}
\end{picture}
\hspace{5mm}
\begin{picture}(200,200)(0,0)
  \LongArrow(15,15)(185,15)
  \Text(195,15)[]{$\kT$} 
  \LongArrow(15,15)(15,185) 
  \Text(15,194)[]{$\pT$} 
  \SetWidth{1.5}
  \Line(50,15)(50,180)
  \Text(50,5)[]{$k_0$}
  \Line(100,15)(100,100)
  \Text(100,5)[]{$k_1$}
  \Line(100,100)(180,180)
  \Text(190,190)[]{$\kT = \pT$}
  \Text(32,90)[]{VMD}
  \Text(75,100)[]{GVMD}
  \Text(140,60)[]{direct}
\end{picture} 
\hspace{5mm}
\begin{picture}(200,200)(0,0)
  \LongArrow(15,15)(185,15)
  \Text(195,15)[]{$\kT$} 
  \LongArrow(15,15)(15,185) 
  \Text(15,194)[]{$\pT$} 
  \SetWidth{1.5}
  \Line(15,15)(180,180)
  \Line(100,15)(100,180)
  \Text(100,5)[]{$Q$}
  \Text(190,190)[]{$\kT = \pT$}
  \Text(70,35)[]{LO DIS}
  \Text(150,60)[]{QCDC+BGF}
  \Text(55,100)[]{non-DGLAP}
  \Text(135,170)[]{photoprod}
\end{picture}   
}     
\end{center}
\caption%
{(a) Schematic graph for a hard $\ga^*\p$ process, illustrating
the concept of three different scales. 
(b) The allowed phase space for this process when $Q^2=0$, 
with one subdivision into event classes.
(c) An alternative classification of the phase space in (b),
which better takes into account unitarization effects.
(d) Event classification in the large-$Q^2$ limit.
\label{FigB}}
\end{figure}

The VMD, direct and anomalous classes have so far been considered 
separately. The complete physics picture presumably would provide 
smooth transitions between the various possibilities. To understand 
the relation between the processes, consider the simple graph of 
Fig.~\ref{FigB}a. There two transverse momentum scales, $\kT$ and 
$\pT$, are introduced (we first consider the case $Q^2=0$). 
Here $\kT$ is related to the $\ga \to \q\qbar$ vertex while 
$\pT$ is the hardest QCD $2 \to 2$ subprocess of the ladder between 
the photon and the proton. (Further softer partons in the ladders are 
omitted for clarity.) The allowed phase space can then conveniently be 
represented by a two-dimensional plane, Fig.~\ref{FigB}b. The region 
$\kT < k_0$ corresponds to a small transverse momentum at the 
$\ga \to \q\qbar$ vertex, and thus to VMD processes. For $\kT > k_0$, 
the events are split along the diagonal $\kT = \pT$. If $\kT > \pT$, 
the hard $2 \to 2$ process of Fig.~\ref{FigB}a is $\ga\g \to \q\qbar$, 
and the lower part of the graph is part of the leading log QCD evolution 
of the gluon distribution inside the proton. These events are direct ones. 
If $\pT > \kT$, on the other hand, the hard process is 
$\qbar \q' \to \qbar \q'$, and the $\ga \to \q\qbar$ vertex builds 
up the quark distribution inside a photon. These events are thus 
anomalous ones.

What complicates the picture is that an event may contain several
interactions, once one considers an incoming particle
as a composite object with several partons that may interact,
more or less independently of each other, with partons from the
other incoming particle. Such a multiple parton--parton interaction 
scenario is familiar already from $\p\p$ physics \cite{multint}.
Here the jet cross section, above some $\pTmin$ scale of the order 
of 2~GeV, increases faster with energy than the total cross section.
Above an energy of a few hundred GeV the calculated jet cross section 
is larger than the observed total one. Multiple interactions offers a 
solution to this apparent paradox, by squeezing a larger number of 
jet pairs into the average event, a process called unitarization 
or eikonalization. The perturbative jet cross section
can then be preserved, at least down to $\pTmin$, but in
the reinterpreted inclusive sense. At the same time, the unitarization
plays a crucial r\^ole in taming the growth of the total 
cross section. 

The composite nature of hadrons also fills another function: it
regularizes the singularity of perturbative cross sections, such as 
$\q\g \to \q\g$, in the limit $\pT \to 0$. Perturbative calculations 
assume free colour charges in the initial and final states of the 
process, while the confinement in hadrons introduces some typical colour
neutralization distance. It is the inverse of this scale that appears
as some effective cutoff scale $\pTmin \simeq 2$~GeV, most likely
with a slow energy dependence \cite{Johann}. One possible 
parameterization is
\begin{equation}
\pTmin(s) = (1.9~{\mr{GeV}}) \left(
\frac{s}{1~\mr{TeV}^2} \right)^{\epsilon} ~,
\label{eq:pTminsdep}
\end{equation}
with the same $\epsilon$ as in eq.~(\ref{sigmatotAB}), since
the rise of the total cross section with energy via Regge 
theory is related to the small-$x$ behaviour of parton distributions
and thus to the density of partons.

Now, if an event contains interactions at several different $\pT$
scales, standard practice is to classify this event by its hardest
interaction. 
With this prescription, the cross section for an event of scale $\pT$ 
is the naive jet cross section at this $\pT$ scale \textit{times} 
the probability that the event contained no interaction at a scale
above $\pT$. The latter defines a form factor, related to
probability conservation.
At large $\pT$ values
the probability of having an even larger $\pT$ is small, i.e.
the form factor is close to unity, and the perturbative cross 
section is directly preserved in the event rate. At lower
$\pT$ values, the likelihood of a larger $\pT$ is increased,
i.e. the form factor becomes smaller than unity, and the rate of 
events classified by this $\pT$ scale falls below the perturbative
answer. 

We expect this picture to hold also for the VMD part of the photon,
since this is clearly in the domain of hadronic physics. Thus,
in the VMD domain $\kT < k_0$, the region of large $\pT$ in 
Fig.~\ref{FigB}b is populated according to perturbation theory,
though with nonperturbative input to the parton distributions.
The region of smaller $\pT$ is suppressed, since the form 
factor here drops significantly below unity. 

As one moves away from the ``pure'' VMD states, such as the $\rho^0$,
much of the same picture could well hold. Interactions at a
larger $\kT$ value could be described in terms of some $\rho'$ state.
The uncertainty relation gives us that a state of virtuality 
$\simeq \kT$ has a maximal size $\simeq 1 / \kT$ and thus spans an 
area $\propto 1 / \kT^2$. 
Such a state could undergo one or several interactions of the 
anomalous-type or remain as a ``low-$\pT$'' direct event.
It is then reasonable to assume that the unitarized cross section 
is proportional to the area of the state interacting
with the proton, i.e. a (kind of) geometrical scaling. 
The colour neutralization distance inside a more virtual photon
state is also reduced, so that the interactions in general tend to 
be weakened by interference effects not included in the simple 
perturbative cross sections. This could then be the origin for
a geometrical scaling like the one in eq.~(\ref{eq:anoint}). 

Calculating the perturbative anomalous cross section in the region
$\pT > \max(\kT, \pTmin(s))$, the geometric scaling answer is exceeded
for some region $\kT \lessim k_1$, with $k_1 \approx 2-4$~GeV
(higher for higher energies). Only for $\kT > k_1$ is the jet cross 
section dropping below the geometric scaling one. At these larger 
$\kT$ values, the direct rate dominates over the anomalous. As a 
convenient but rather arbitrary choice, for subsequent studies we put 
$k_1 = \pTmin(s)$, with the latter given by eq.~(\ref{eq:pTminsdep}).

The final scenario is illustrated in Fig.~\ref{FigB}c.
The bulk of the cross section, in the region $\kT < k_1$,
is now described by the photon interacting as dense, hadronic states,
VMD for $\kT \lessim k_0$ and Generalized VMD (GVMD) for
$k_0 \lessim \kT \lessim k_1$. The total VMD cross section is 
given by the pomeron-type ansatz, while the jet cross section
can be obtained from the parton distributions of the respective
vector meson state. Correspondingly, the GVMD states have a
total cross section based on Pomeron considerations and a jet
cross section now based on the anomalous part of the parton
distributions of the photon. In principle, an eikonalization
should be performed for each GVMD state separately, but in 
practice that would be overkill. Instead the whole region is
represented by one single state per quark flavour, with a jet 
production given by the full anomalous part of the photon 
distributions. 

Thus, post facto, the approximate validity of a Regge theory ansatz 
for $\sigma_{\mr{tot}}^{\ga\p}$ is making sense. Above $k_1$
only the direct cross section need be considered, since here the
anomalous one is negligibly small, at least in terms of total cross 
sections. (As noted above, we have actually chosen to lump it with
the other GVMD contributions, so as not to lose the jet rate itself.)

\subsection{Deeply Inelastic Scattering}

At not too large $Q^2$, the Deeply Inelastic Scattering of a 
high-energy charged lepton off a proton target, involves a single photon 
exchange between a beam lepton and a target quark. 
The double-differential $\e\p \rightarrow \e\X$ cross-section 
for DIS can be expressed in terms of the total cross-section 
for virtual transverse (T) and longitudinal (L) photons \cite{sigmaTL}: 
$
\d^2\sigma/\d y/\d Q^2 =  
f_{\ga/\e}^{\mr{T}}(y,Q^2) \sigma_\mr{T}(y,Q^2) + 
f_{\ga/\e}^{\mr{L}}(y,Q^2) \sigma_\mr{L}(y,Q^2) ~,
$
where $f_{\ga/\e}^{\mr{T,L}}$ are the transverse and longitudinal fluxes 
\cite{lutp9911}.

The total virtual photon-proton cross section can be related to the 
proton structure function $F_2$ by~\cite{F2FLdef}
\begin{equation}
\sigma_\mr{tot}^{\gast\p}
\equiv\sigma_\mr{T}+\sigma_\mr{L}
\simeq\frac{4\pi^2\aem}{Q^2} F_2(x,Q^2)
=\frac{4\pi^2\aem}{Q^2} 
\sum_{\q} e_{\q}^2 \, \left\{ x  q(x, Q^2) + 
x \overline{q}(x,Q^2) \right\}\;.
\label{eq:F2}
\end{equation} 
where the last equality is valid for the parton model to lowest order. 
Such an interpretation is not valid in the limit 
$Q^2 \to 0$, where gauge invariance requires $F_2(x, Q^2) \to 0$
so that $\sigma_\mr{tot}^{\gast\p}$ remains finite. We will 
replace the DIS description by a photoproduction one in this limit. 
Hence, at small photon virtualities, the DIS process $\gast \q \ra \q$ 
should be constructed vanishingly small as compared to the contribution 
from the interaction of the hadronic component of the photon. To obtain 
a well-behaved DIS cross 
section in this limit, a $Q^4/(Q^2+m_\rho^2)^2$ factor is introduced. 
Here $m_\rho$ is some non-perturbative hadronic parameter, for 
simplicity identified with the $\rho$ mass. Then, in the parton model,
eq.~(\ref{eq:F2}) modifies to a DIS cross section
\begin{equation}
\sigma_\mr{DIS}^{\gast\p}
\simeq\frac{4\pi^2\aem Q^2}{(Q^2+m_\rho^2)^2} \sum_{\q} e_{\q}^2 \, 
\left\{ x  q(x, Q^2) + x \overline{q}(x,Q^2) \right\}\;.
\label{eq:F2mod}
\end{equation} 
For numerical studies, the available parton distribution parameterizations 
for the proton have some lower limit of applicability in both $x$ and $Q^2$. 
For values below these minimal ones, the parton distributions are frozen at 
the lower limits.

In DIS, the photon virtuality $Q^2$ 
introduces a further scale to the process in Fig.~\ref{FigB}a.
The traditional DIS region is the strongly ordered one,
$Q^2 \gg \kT^2 \gg \pT^2$, where DGLAP-style evolution \cite{DGLAP}
is responsible for the event structure. As above, ideology 
wants strong ordering, while real life normally is
based on ordinary ordering $Q^2 > \kT^2 > \pT^2$.
Then the parton-model description 
of $F_2(x,Q^2)$ in eq.~(\ref{eq:F2}) is a very good first 
approximation. The problems come when the ordering is no longer 
well-defined, i.e. either when the process contains several large 
scales or when $Q^2 \to 0$. In these regions, an $F_2(x,Q^2)$ may
still be defined by eq.~(\ref{eq:F2}), but its physics interpretation 
is not obvious.
 
Let us first consider a large $Q^2$, where a possible classification
is illustrated in Fig.~\ref{FigB}d. The regions $Q^2 > \pT^2 > \kT^2$ 
and $\pT^2 > Q^2 > \kT^2$ correspond to non-ordered emissions, that 
then go beyond DGLAP validity and instead have to be described by the 
BFKL \cite{BFKL} or CCFM \cite{CCFM} equations, see e.g. \cite{LDC}. 
Normally one expects such cross sections to be small at large $Q^2$. 
The (sparsely populated) region $\pT^2 > \kT^2 > Q^2$ can be viewed 
as the interactions of a resolved (anomalous) photon.

The region $\kT^2 > Q^2 \gg 0$ and $\kT^2 > \pT^2$ contains the 
${\mathcal O}(\as)$ corrections to the lowest-order (LO) DIS process 
$\ga^* \q \to \q$, namely QCD Compton $\ga^* \q \to \q \g$ 
and boson-gluon fusion $\ga^* \g \to \q \qbar$. These 
are nothing but the direct processes  $\ga \q \to \q \g$ and 
$\ga \g \to \q \qbar$ extended to virtual photons. The borderline
$\kT^2 > Q^2$ is here arbitrary --- also processes with $\kT^2 < Q^2$
could be described in this language. In the parton model, this whole 
class of events are implicitly included in $F_2$, and are related
to the logarithmic scaling violations of the parton distributions.
The main advantage of a separation at $\kT = Q$ thus comes from the 
matching to photoproduction. Also the exclusive modeling of events, 
with the attaching of parton showers of scale $Q^2$ to DIS events,
is then fairly natural.

The DIS cross section thus is subdivided into
$
\sigma_\mr{tot}^{\gast\p}
 \simeq  \sigma_{\mr{DIS}}^{\gast\p} =
\sigma_{\mr{LO\,DIS}}^{\gast\p} +
\sigma_{\mr{QCDC}}^{\gast\p} + \sigma_{\mr{BGF}}^{\gast\p}
\;.
$
The $\sigma_{\mr{DIS}}^{\gast\p}$ is given by eq.~(\ref{eq:F2mod}),
while the last two terms are well-defined by an integration of the 
respective matrix element \cite{lutp9911}. When extended to small $Q^2$, 
these two terms will increase in importance, and one may eventually 
encounter an $\sigma_{\mr{LO\,DIS}}^{\gast\p} < 0$, if calculated
by a subtraction of the QCDC and BGF terms from the total DIS cross 
section. However, here we expect the correct answer not to be a 
negative number but an exponentially suppressed one, by a Sudakov form 
factor. This modifies the cross section: 
\begin{equation}
\sigma_{\mr{LO\,DIS}}^{\gast\p} = \sigma_{\mr{DIS}}^{\gast\p} -
\sigma_{\mr{QCDC}}^{\gast\p} - \sigma_{\mr{BGF}}^{\gast\p}
~~ \longrightarrow ~~ 
\sigma_{\mr{DIS}}^{\gast\p} \; \exp \left( - \frac{%
\sigma_{\mr{QCDC}}^{\gast\p} + \sigma_{\mr{BGF}}^{\gast\p}}%
{\sigma_{\mr{DIS}}^{\gast\p}} \right) \;.
\label{eq:LODISmod}
\end{equation}
Since we here are in a region where
$\sigma_{\mr{LO\,DIS}}^{\gast\p} \ll \sigma_\mr{DIS}^{\gast\p}$,
i.e. where the DIS cross section is no longer the dominant one, this
change of the total DIS cross section is not essential. Even more,
for $Q^2 \to 0$ we know that the direct processes should survive whereas 
the lowest-order DIS one has to vanish. Since eq.~(\ref{eq:F2mod})
ensures that $\sigma_{\mr{DIS}}^{\gast\p} \to 0$  in this limit,
it also follows that $\sigma_{\mr{LO\,DIS}}^{\gast\p}$ does so.   

\section{From Real to Virtual Photons}

It is now time to try to combine the different aspects of the photon, 
to provide an answer that smoothly interpolates between the 
photoproduction and DIS descriptions, in a physically sensible way.

A virtual photon has a reduced probability to fluctuate into a vector 
meson state, and this state has a reduced interaction probability. 
This can be modeled with the traditional dipole factors 
\cite{dipolevirt} introduced to eq.~(\ref{sigmatotVMDp}).
Similarly, the GVMD states are affected, where a relation 
$2 \kT \simeq m$ is assumed.

The above generalization to virtual photons does not address the issue 
of longitudinal photons. Their interactions vanish in the limit 
$Q^2 \to 0$, but can well give a non-negligible contribution at
finite $Q^2$ \cite{longitdata}. A common approach is to attribute the 
longitudinal cross section with an extra factor of $r_V=a_V Q^2/m_V^2$ 
relative to the transverse one \cite{longitmod}, where $a_V$ is some 
unknown parameter to be determined from data. Such an ansatz only 
appears reasonable for moderately small $Q^2$, however, so following 
the lines of our previous study of jet production by virtual 
photons~\cite{lutp9911}, we will try the two alternatives 
\begin{eqnarray}
r_1(m_V^2, Q^2) = 
a \frac{m_V^2 Q^2}{(m_V^2 + Q^2)^2} \qquad
r_2(m_V^2, Q^2) = 
a \frac{Q^2}{(m_V^2 + Q^2)}
\end{eqnarray}
While $r_1$ vanishes for high $Q^2$, $r_2$ approaches the constant 
value $a$. The above VMD expressions are again extended to GVMD by the 
identification $m_V \approx 2\kT$. The GVMD cross section can then be 
written as
\begin{eqnarray}
\sigma_{\mr{GVMD}}^{\gast\p} = 
\frac{\aem}{2\pi}  \sum_\q 2 \e_\q^2 \int_{k_0^2}^{k_1^2} 
\frac{\d \kT^2}{\kT^2} 
\left[ 1+r_i(4 \kT^2, Q^2)\right]
\left( \frac{4 \kT^2}{4 \kT^2+Q^2} \right)^2 
\frac{k_{V(\q\qbar)}^2}{\kT^2} 
\sigma^{V(\q\qbar)\p}(W^2) \;.
\label{eq:sigmatotAnopTL}
\end{eqnarray}

The extrapolation to $Q^2 > 0$ is trivial for the direct processes,
which coincide with the DIS QCDC and BGF processes. The matrix elements
contain all the required $Q^2$ dependence, with a smooth behaviour in
the $Q^2 \to 0$ limit. They are to be applied to the region
$\kT > \max(k_1, Q)$ (and $\kT > \pT$, as usual). 

Remains the LO DIS process. It is here that one could encounter an
overlap and thereby double-counting with the VMD and GVMD processes.
Comparing Fig.~\ref{FigB}d with Fig.~\ref{FigB}c, one may note that
the region $\pT > \kT$ involves no problems, since we have made no
attempt at a non-DGLAP DIS description but cover this region entirely
by the VMD/GVMD descriptions. Also, if $Q > k_1$, then the region 
$k_1 < \kT < Q$ (and $\kT > \pT$) is covered by the DIS process only.
So it is in the corner $\kT < k_1$ that the overlap can occur. 
If $Q^2$ is very small, the exponential factor in 
eq.~(\ref{eq:LODISmod}) makes the DIS contribution too small to worry 
about. Correspondingly, if $Q^2$ is very big, the VMD/GVMD contributions
are too small to worry about. Furthermore, a large $Q^2$ implies
a Sudakov factor suppression of a small $\kT$ in the DIS description. 
If $W^2$ is large, the multiple-interaction discussions above are 
relevant for the VMD/GVMD states: the likelihood of an interaction at 
large $\pT$ will preempt the population of the low-$\pT$ region. 

In summary, it is only in the region of intermediate $Q^2$ and rather
small $W^2$ that we have reason to worry about a significant
double-counting. Typically, this is the region where
$x \approx Q^2/(Q^2 + W^2)$ is not close to zero, and where $F_2$ is
dominated by the valence-quark contribution. The latter behaves roughly
$\propto (1-x)^n$, with an $n$ of the order of 3 or 4. Therefore
we will introduce a corresponding damping factor to the VMD/GVMD terms.
The real damping might be somewhat different but, since small $W$ values
are not our prime interest, we rest content with this approximate form.

In total, we have now arrived at our ansatz for all $Q^2$
\begin{equation}
\sigma_\mr{tot}^{\gast\p}  = 
\sigma_{\mr{DIS}}^{\gast\p} \; \exp \left( - \frac{%
\sigma_{\mr{dir}}^{\gast\p}}{\sigma_{\mr{DIS}}^{\gast\p}} \right) +
\sigma_{\mr{dir}}^{\gast\p} +
\left( \frac{W^2}{Q^2 + W^2} \right)^n 
\sigma_{\mr{res}}^{\gast\p} \;.
\label{eq:siggastptotsimple}
\end{equation}
To keep the terminology reasonably compact, also for the 
$\gast\gast$ case below, we use res as shorthand for the 
resolved VMD plus GVMD contributions and dir as shorthand for the 
QCDC and BGF processes. The DIS and GVMD terms are given by 
eqs.~(\ref{eq:F2mod}) and (\ref{eq:sigmatotAnopTL}), respectively,
and the QCDC and BGF terms by direct integration of the respective
matrix elements for the region $\kT > \max(k_1, Q)$. 
(The VMD contribution is obtained from eq.~(\ref{sigmatotVMDp}) 
with the appropriate dipole and polarization factors applied.)

The extension to $\gast\gast$ follows from the
$\gast\p$ formalism above, but now with (up to) five scales to keep 
track of: $\pT$, $\kTone$, $\kTtwo$, $Q_1$ and $Q_2$. First consider
the three by three classes present already for real photons, which
remain nine distinct ones for $Q_1^2 \neq Q_2^2$. Each VMD or GVMD
state is associated with its dipole damping factor and its correction
factor for the longitudinal contribution. The QCDC and BGF matrix
elements involving one direct photon on a VMD or a GVMD state explicitly
contain the dependence on the direct photon virtuality, separately
given for the transverse and the longitudinal contributions. Also the
direct$\times$direct matrix elements are known for the four possible
transverse/longitudinal combinations. 

To this should be added the new DIS processes that appear for
non-vanishing $Q^2$, when one photon is direct and the other
resolved, i.e. VMD or GVMD. For simplicity, first assume that one 
of the two photons is real, $Q_2^2=0$. For large $Q_1^2$, this 
DIS contribution $\sigma_{\mr{DIS}\times\mr{res}}^{\gast\ga}(Q_1^2)$
can be given a parton-model interpretation similarly to eq.~(\ref{eq:F2})
and (\ref{eq:F2mod}).
Note that this is only the resolved part of 
$\sigma_{\mr{tot}}^{\gast\ga}$. The direct contribution from
$\gast\ga \to \q\qbar$ comes in addition, but can be neglected
in the leading-order definition of $F_2^{\ga}$. We will therefore
use parton distribution parameterizations for the resolved photon, 
like SaS 1D \cite{saspdf}, to define the
$\sigma_{\mr{DIS}\times\mr{res}}^{\gast\ga}(Q_1^2)$.
Then eq.~(\ref{eq:siggastptotsimple}) generalizes to
\begin{eqnarray}
\sigma_{\mr{tot}}^{\gast\ga} =
\sigma_{\mr{DIS}\times\mr{res}}^{\gast\ga}  
\exp \left( - \frac{\sigma_{\mr{dir}\times\mr{res}}^{\gast\ga}}%
{\sigma_{\mr{DIS}\times\mr{res}}^{\gast\ga}} \right) +
\sigma_{\mr{dir}\times\mr{res}}^{\gast\ga} 
+ \sigma_{\mr{res}\times\mr{dir}}^{\gast\ga} +
\sigma_{\mr{dir}\times\mr{dir}}^{\gast\ga} +
\left( \frac{W^2}{Q_1^2 + W^2} \right)^n 
\sigma_{\mr{res}\times\mr{res}}^{\gast\ga}  \;.
\label{eq:siggastgatot}
\end{eqnarray}
The large-$x$ behaviour of a resolved photon does not
agree with that of the proton, but for simplicity we will
stay with $n=3$.

Finally, the generalization to both photons virtual gives an extra 
term $\sigma_{\mr{res}\times\mr{DIS}}^{\gast\gast}$ times its 
corresponding exponential factor as in eq.~(\ref{eq:LODISmod}) and 
eq.~(\ref{eq:siggastgatot}). When $Q_1^2 \gg Q_2^2$ the expression for 
$\sigma_\mr{tot}^{\gast\gast} (W^2, Q_1^2, Q_2^2)$
can be related to the structure function of a virtual photon,
$F_2^{\gast}(x,Q^2=Q_1^2,P^2=Q_2^2)$, where 
$x = Q_1^2/(Q_1^2 + Q_2^2 + W^2)$.

\section{Summary and Outlook}

A brief description of a model for real and virtual photons has been 
presented. We have developed one specific approach, where the main 
idea is to classify events by the hardest scale involved. 
This is not an economical route, since it leads 
to many event classes. For studies e.g. of the total cross section in 
the intermediate-$Q^2$ region, it is cumbersome and not necessarily
better than existing approaches \cite{existingsigmagap}. However, by 
the breakdown into distinct event classes, the road is open to provide 
a (more or less) complete picture of all event properties. It is this 
latter aspect that has then guided the model development. 
A more detailed description can be found in a future 
publication~\cite{future}. Some results was presented on the Working group 
meeting in April 2000~\cite{talk}.

\end{document}